\documentclass[12pt]{article}
\usepackage{epsfig}
\begin{document}
\bigskip
\centerline {\bf ELECTRON BEAM M\O LLER POLARIMETER AT JLAB HALL A }
\bigskip
\centerline { A.V. Glamazdin, V.G. Gorbenko, L.G. Levchuk, R.I. Pomatsalyuk, 
A.L. Rubashkin, P.V. Sorokin }
\centerline {\em Kharkov Institute of Physics and Technology, Kharkov, 310108, 
Ukraine}
\centerline { D.S. Dale, B.Doyle, T. Gorringe, W. Korsch, V. Zeps }
\centerline {\em University of Kentucky, Department of Physics and Astronomy, 
Lexington, KY 40506-0055, USA}
\centerline { J.P.Chen, E. Chudakov, S. Nanda, A. Saha }
\centerline {\em Thomas Jefferson National Accelerator Facility, Newport 
News, 23606-4350, VA, USA }
\centerline { A. Gasparian }
\centerline {\em Hampton University, Hampton, Department of Physics, VA, 23668, 
USA }

\bigskip
\centerline {\bf ABSTRACT }
\bigskip

As  part of the spin physics program at Jefferson Laboratory (JLab), a M\o ller 
polarimeter was developed to measure the polarization of electron beam of energies $0.8$ 
to $5.0$ $GeV$. A unique signature for M\o ller scattering is obtained using a 
series of three quadrupole magnets which provide an angular selection, and a dipole 
magnet for energy analysis. The design, commissioning, and the first results of the 
polarization measurements of this polarimeter will be presented as well as future 
plans to use its small scattering angle capabilities to investigate physics in the very 
low $Q^2$ regime.
\bigskip

M\o ller polarimeters are widely used for electron beam polarization measurements 
in the $GeV$ energy range. The high quality of polarization experiments anticipated 
at new-generation CW multi-GeV electron accelerators, such as Jefferson Laboratory 
(JLab), require precise measurements of electron beam parameters. One of the 
parameters is the electron beam polarization. The Hall A beam line at JLab is equipped 
with a M\o ller polarimeter. It was designed and constructed in collaboration with 
Jefferson Laboratory, the Kharkov Institute of Physics and Technology and the 
University of Kentucky.

\begin{figure}[ht]
\vskip 0.1cm
\epsfig{file=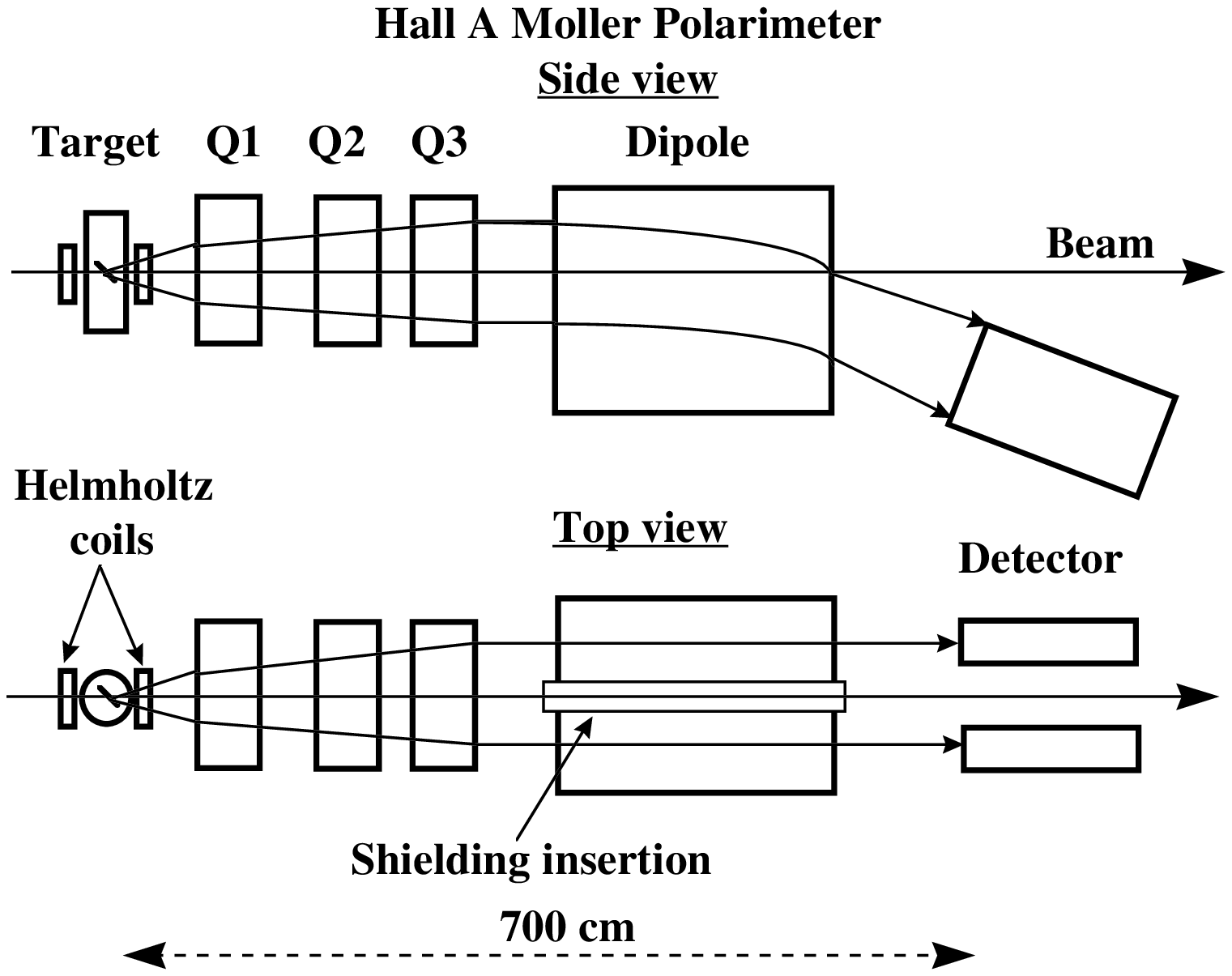, width=16cm, height=8cm}
\large{ Fig. 1.} 
\large{ Hall A M\o ller polarimeter set-up}
\end{figure}

The polarimeter is schematically presented in Fig. 1. The horizontal plane in Hall A 
is in accordance to the polarimeter reaction plane. The polarimeter consists of a 
polarized electron target, three quadrupole magnets, a dipole magnet and a detector. 
The polarimeter quadrupole magnets make it possible to keep the position of all 
polarimeter elements unchanged within the whole range of JLab energies. Their 
primary purpose is to focus the divergent trajectories of M\o ller electrons in the 
scattering plane into a paired trajectories aligned with the axis of the beam at the 
exit of the last quadrupole. The dipole is the main element of the polarimeter magnetic 
system. It provides the energy analysis, thus separating the M\o ller scattered 
electrons ($E_{o}/2$, $\Theta_{Moll}$) from electrons coming from the Mott scattering 
peak ($E \sim E_{o}$, $\Theta \sim \Theta_{Moll}$) and thereby suppressing the 
background. It also bends the M\o ller electrons from the reaction plane, allowing 
their detection away from the electron beam. The dipole has a magnetic shielding 
insertion in the center of the magnetic gap. The M\o ller electrons pass through the 
dipole on the left and right sides of this shielding insertion. The primary electron 
beam passes through a $4$ $cm$ diameter hole  bored in the shielding insertion letting 
its passage to the Hall A beam dump with small influence of the dipole magnetic field.

The M\o ller polarimeter detector is located in the shielding box downstream of the 
dipole and consists of two modules (left and right) for coincidence measurements. 
Each part of the detector includes an aperture detector made of plastic scintillator 
and four blocks of lead glass. 

The polarized electron target chamber contains two target frames with different 
ferromagnetic foils: 
\begin{enumerate}
\item $99.95\%$ pure iron foil $10.9$ $\mu m$ thick with effective polarization $7.12\%$ 
in an applied magnetic field of about $28$ $mT$;
\item foil of Supermendur $13.9$ $\mu m$ thick with an effective polarization $7.6\%$ 
in an applied magnetic field of about $28$ $mT$.
\end{enumerate}

The targets are mounted on a vertical ladder and are cooled with liquid nitrogen 
down to $115$ $K$. They can be rotated in an angular range $\pm (20^{o}- 160^{o})$. 
The beam-polarization measurements are made with one or the other foil in the 
beam ({\em TOP} or {\em BOTTOM} target position). The third position ({\em HOME}) 
is used when beam polarization measurement is not in progress. The target foil 
magnetization is measured by a series of pickup coils. The first polarization measurement 
with the Hall A M\o ller polarimeter was done in June 1997, regular measurements with 
the polarimeter are made. Results of a measurement are shown in Fig. 2. 

\begin{figure}[ht]
\epsfig{file=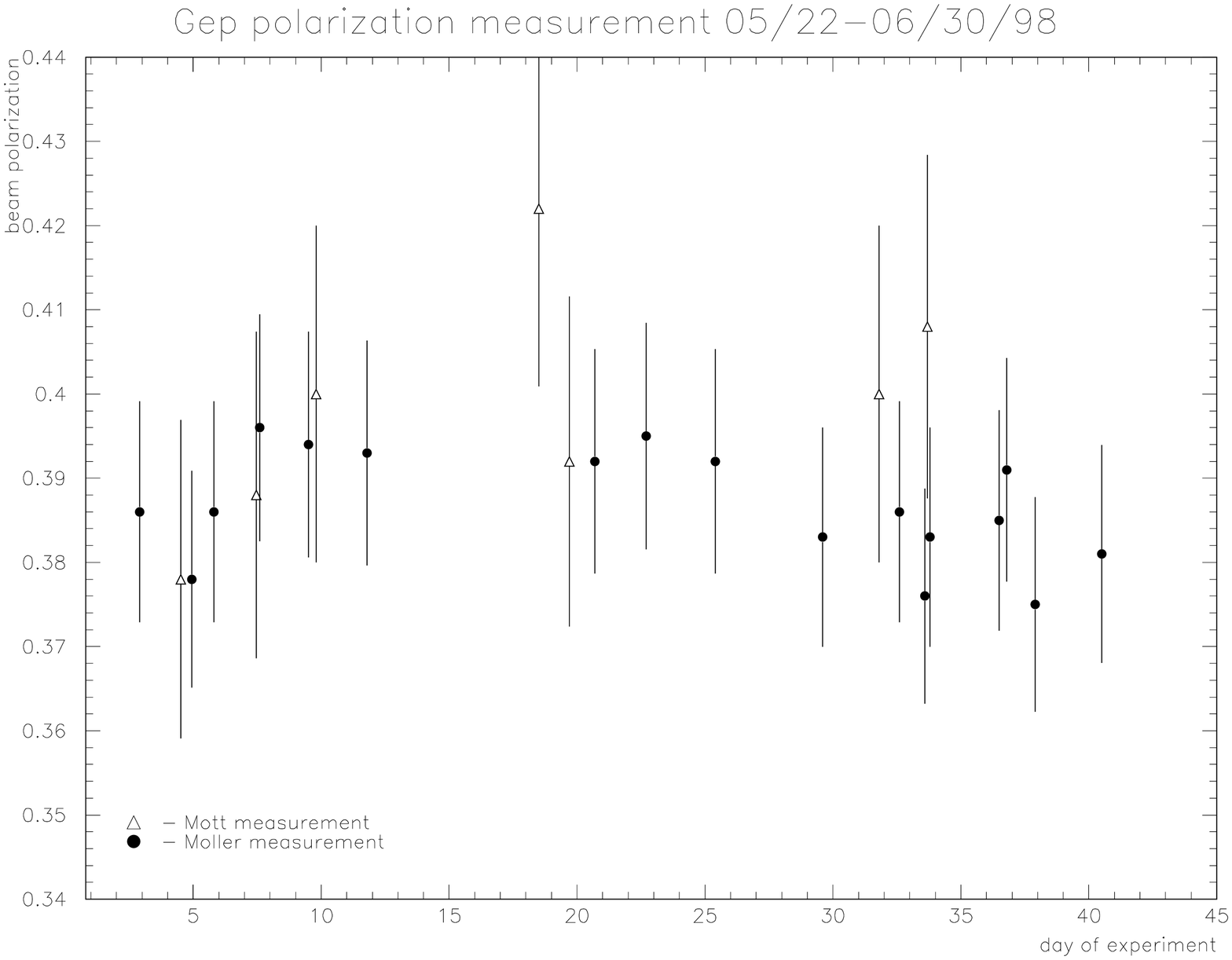, width=14cm, height=8cm}
\vskip 0.1cm
\large{ Fig. 2.} 
\large{ Result of the Hall A electron beam polarization measurements 
with M\o ller and Mott polarimeters for experiment E-93-027. }
\end{figure}

The Hall A M\o ller polarimeter covers the energy range $0.8$ to $5.0$ $GeV$ and 
can be used for measurements with beam currents from $0.5 - 5.0$ $\mu A$. About 
twenty minutes of measurement time is needed to take data with a statistical error 
of less than  $1\%$. Although the polarimeter quadrupole magnets are part of the 
regular Hall A beam transport, it is not necessary to change the  quadrupole magnet 
settings or the primary beam trajectory in switching from data taking with the 
Hall A physics target to a polarization measurement. Also, the polarization 
measurements can be done with the Hall A fast raster on.

A typical beam current for the polarization measurement is $1.5$ $\mu A$ when the
iron target heats up to $150$ $K$. This target heating provides a relative target 
depolarization of $0.3\%$. The Helmholtz coils provide a $28$ $mT$ magnetic field 
in the area in which the electron beam passes through the target. The experimentally 
measured target polarization is $7.6\%$ for the Supermendur foil and $7.12\%$ for the 
iron foil. The background in coincidence measurements is negligible. The typical 
detector acceptance in the reaction plane for an energy range $2 - 5$ $GeV$ is $\Delta 
\Theta_{Moll}\sim \pm 14\raisebox{1ex}{\scriptsize o}$ in c.m. and is about $76 \%$ 
in analyzing power. The Levchuk effect $\cite{levchuk}$, $\cite{afanas}$  is estimated 
to be about $2$ $\%$ and was not observed at the $\sim3$ $\%$ level of the systematic 
error of our measurements. Other sources of systematic errors were considered and 
are summarized in Table 1.

\vspace*{0.5cm}
\begin{tabular}{| l | l  | l | l | r |}  \hline  
Parameter&$ < A_{zz} > $ & $\cos\Theta_{targ}$ & $P_{targ}$ & $ {\bf Total:} $ 
\\ \hline
Error    & $0.25$ $ \% $& $\leq 1$ $\% $ & $\leq 3$ $\%$ & {\bf$\sim3\%$} 
\\ \hline
\end{tabular} 
\vspace*{0.5cm}

In addition to being a part of the standard beam line instrumentation in Hall A, the 
small scattering angle capabilities of the M\o ller polarimeter, coupled with the 
momentum analyzing capabilities of its dipole, present unique opportunities to do 
physics in very low $Q^2$ regime. The $QQQD$  design of the M\o ller spectrometer will 
make possible the  electron scattering experiments at electron scattering angles ranging 
from about three degrees to less than one degree with $\Delta p^{\prime} /p^{\prime}$ 
of about $10^{-3}$. As an initial area of investigation, we intend to measure the 
neutral pion form factor, $F_{\gamma^{\ast}\gamma\pi^{o}}$, at low $Q^2$ via the 
virtual Primakoff effect $\cite{e97009}$, {\em i.e.} $\pi^o$ electroproduction in 
the Coulomb field of  heavy nucleus. The slope of this form factor in the low 
$Q^2$ range to be measured, $0.005$ $(GeV/c)^2$ to $0.04$ $(GeV/c)^2$, gives a 
measure of the mean square $\gamma^{\ast}\gamma\pi^o$ interaction radius and is 
sensitive to the constituent quark mass. Such an experiment can be performed by 
removing the third quadrupole magnet, installing position sensitive detectors in the 
focal plane, and placing a series of lead glass photon detectors upstream of the 
dipole to measure the $\pi^o$ decay photons from the $Pb(e,e^{\prime}\pi^o)Pb$ reaction.

A double arm M\o ller polarimeter, used to measure the polarization of a  $0.8-5.0$ $GeV$ 
primary electron beam in Hall A of JLab, has been described. It is used for the extensive 
planned spin physics program in Hall A. The polarimeter has been found to be robust and 
stable. Statistical errors of between $0.2$ and $0.8\%$ per measurement have made 
precision tests of possible systematic shifts in the data possible. Combining the 
systematic uncertainties leads to a final determination of the beam polarization with 
a relative uncertainty of $\leq3\%$.  More detailed information about the polarimeter 
design, status and current measurements is available at  http:$\backslash\backslash$www.
jlab.org$\backslash$\~ \ moller.

\smallskip
\centerline{\em Acknowledgments}
\smallskip

This research was supported in part by Ukrainian Ministry of Science and Technology 
under Contract No. F5/1758-97 project No. 2.5.1/27.

\end{document}